\documentclass{article}
\usepackage{spconf,amsmath,graphicx,algorithm,algorithmic,amssymb,booktabs,multirow}

\usepackage{enumitem}
\setlist{nosep, leftmargin=14pt}

\usepackage{mwe} % to get dummy images

\title{LMPDnet: TOF-PET list-mode image reconstruction using model-based deep learning method}
\name{Chenxu Li \qquad Rui Hu \qquad Jianan Cui \qquad Huafeng Liu}

\address{
	State Key Lab of Modern Optical Instrumentation \\
	Zhejiang University \\
	Hangzhou, 310027, China }

\begin{document}
%\ninept
%
\maketitle
\begin{abstract}
The integration of Time-of-Flight (TOF) information in the reconstruction process of Positron Emission Tomography (PET) yields improved image properties. However, implementing the cutting-edge model-based deep learning methods for TOF-PET reconstruction is challenging due to the substantial memory requirements. In this study, we present a novel model-based deep learning approach, LMPDNet, for TOF-PET reconstruction from list-mode data. We address the issue of real-time parallel computation of the projection matrix for list-mode data, and propose an iterative model-based module that utilizes a dedicated network model for list-mode data. Our experimental results indicate that the proposed LMPDNet outperforms traditional iteration-based TOF-PET list-mode reconstruction algorithms. Additionally, we compare the spatial and temporal consumption of list-mode data and sinogram data in model-based deep learning methods, demonstrating the superiority of list-mode data in model-based TOF-PET reconstruction.
\end{abstract}
\section{Introduction}
\label{sec:intro}

Incorporating Time-of-Flight (TOF) information in Positron Emission Tomography (PET) imaging enhances image quality and bolsters lesion detectability\cite{karp2008benefit,surti2020update}. The main methods for TOF-PET reconstruction include Ordered Subset Expectation Maximization (OSEM)\cite{snyder1983image,pratx2010fast}, Maximum A Posteriori (MAP)\cite{bai2014map}, Primal-Dual Hybrid Gradient (PDHG)\cite{schramm2022fast}, etc. These algorithms are suitable for both sinogram data and list-mode data. However, these iterative-based algorithms suffer from computational inefficiency and may produce reconstructed images with very poor quality when applied to the data of low counts.

In recent times, machine learning has become a widely-utilized technique used in the field of medical image reconstruction due to its high computational efficiency and data-driven characteristic. Numerous studies have demonstrated the efficacy of model-based deep learning methods in reconstructing high-quality PET images\cite{mehranian2020model,reader2020deep,kim2018penalized}, and the model-based approach also provides interpretability to deep learning\cite{hyun2021deep}. However, there are few model-based deep learning methods for TOF-PET reconstruction. On one hand, the storage requirements of the system matrix corresponding to the sinogram combined with TOF information will be even greater. On the other hand, starting from list-mode data, how to integrate the system matrix of list-mode data into the iterative framework is a problem worth discussing. For a 2D image with a dimension of 128$\times$128 and sinogram with a dimension of 357$\times$224$\times$17 (rad$\times$view$\times$tofbin), the system matrix occupies $\sim$83GB of video memory. What’s worse, as the increase of axial field of view, its memory usage will also increase dramatically. This severely limits the application of model-based learning methods on TOF-PET reconstruction. The reconstruction algorithm based on list-mode data can discard the fixed format like sinogram. It ensures that the data are all valid information, rather than a large number of ``empty bins" appearing in the sinogram in TOF-PET, avoiding the waste of memory. The memory occupied by the list-mode data depends on the number of coincident events detected in one scan, regardless of whether there is TOF information or not. 
For the example mentioned above, if the count of coincident events is 1e5, the projection matrix (corresponding to the system matrix) occupies only $\sim$6GB of video memory.

In this work, we proposed the LMPDNet, i.e. List-mode Primal Dual Net, a novel model-based deep learning approach for list-mode data to improve the TOF-PET reconstruction. We first address the issue of real-time calculation of the projection matrix during iterative training. Based on the projection matrix calculation method proposed by Joseph\cite{joseph1982improved}, we used CUDA to compute the projection of each response line (LOR) in parallel to achieve acceleration. Then, based on the Learned Primal-Dual method\cite{adler2018learned, guazzo2021learned}, the LMPDNet was proposed. 
Simulation experiments showed that our method was superior to the current mainstream TOF-PET list-mode reconstruction method\cite{pratx2010fast, schramm2022fast}. To the best of our knowledge, this is the first unrolled model-based deep learning method for TOF-PET list-mode reconstruction.

\begin{figure*}[t]
    \centering
    \includegraphics[width=17.5cm]{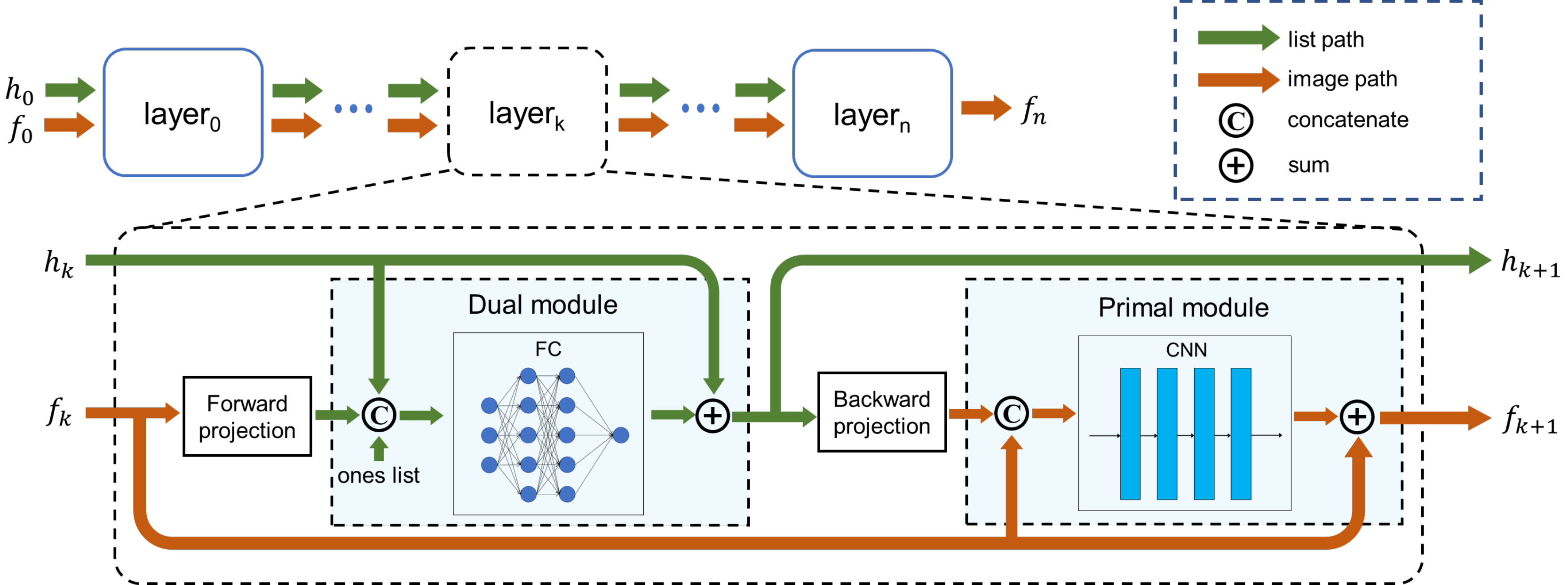}
    \hfill
    \caption{Overall framework of LMPDNet}
    \label{fig:network_structure}
\end{figure*}

\section{METHOD}
\label{sec:method}

\subsection{Learned Primal-Dual}
\label{ssec:Learned Primal-Dual}

Mathematically, the TOF-PET reconstruction problem from list-mode can be formulated as:
\begin{equation}
\label{equ1}
    g=Pf+noise
\end{equation}
where $f\in X$(image domain) and $g \in Y$(measurement domain), and $P$ is the forward projection matrix that mapping $X \rightarrow Y$ with TOF information.

A general approach to solve (\ref{equ1}) is to minimize the sum of the negative log-likelihood function $\mathcal{L}(Pf,g)$ and penalized regularization term $\mathcal{S}(f)$:
\begin{equation}
\label{equ2}
\underset{f\in X}{min} [ \mathcal{L}(Pf,g)+\lambda\mathcal{S}(f)]
\end{equation}
where the $\lambda$ is the regularization parameter. This is a large-scale optimization problem. A general framework called the Learned Primal-Dual\cite{adler2018learned} combining deep learning with PDHG was proposed for solving optimization problems of the form similar to (\ref{equ2}). PDHG is based on the use of forward and backward projections, and is capable of solving problems with complex constraints, such as positivity and sparsity.
%Shown in Algorithm \ref{alg1}, a learning-based method named Learned Primal-Dual\cite{adler2018learned} is proposed to solve optimization problems of the form similar to (\ref{equ2}). 
%This method replaces the proximal operator in the traditional PDHG algorithm with a parameterized operator, where the parameters are learned from the training data.
Shown in Algorithm \ref{alg1}, The main idea of Learned Primal-Dual is to unroll the PDHG algorithm, keeping the forward projection operator $\mathcal{T}$ and its adjoint projection operator $\mathcal{T}^{*}$, and replacing the proximal operator in the PDHG algorithm with the parameterized operator $\Gamma_{\theta^{d}}$ and $\Lambda_{\theta^{p}}$, \iffalse where $\Gamma_{\theta^{d}$}$ is used to perform data-to-data mapping, $\Lambda_{\theta^{p}}$ is used to perform image-to-image mapping,\fi and the parameters $\theta^{d}$ and $\theta^{p}$ are learned from the training data. $h_{k}$ and $f_{k}$ represents the data in measurement domain and image domain respectively. For list-mode data, the operator $\mathcal{T}$ is the projection matrix $P$.

%In order to apply the Learned Primal-Dual algorithm to list-mode TOF-PET, we have to solve the problem of computing the projection operators $\mathcal{K}$ and $\mathcal{K}^{*}$ in list-mode data, as well as choosing the appropriate neural network model to fit $\Gamma_{\theta^{d}$ and $\Lambda_{\theta^{p}$.

\subsection{Overall Framework}
\label{ssec:Overall Framework}

The overall framework of the LMPDNet is shown in Figure \ref{fig:network_structure}. Like the Learned Primal-Dual reconstruction method, the structure of the proposed method is formed by concatenating multiple identical layers. The inputs are list-mode space input $h_{0}\in \mathbb{R}^{n}$ and image space input $f_{0}\in \mathbb{R}^{W\times H}$ initialized to zero, where $n$ is the length of list-mode data, $W$ and $H$ are width and height of the image. Each layer consists of four parts: Forward projection, Dual module, Backward projection and Primal module.

\begin{algorithm}[t]
  \caption{Learned Primal-Dual}
  \label{alg1}
  \begin{algorithmic}[1]
    \STATE Initialize $f_{0}(=0), h_{0}(=0)$
    \FOR {$k=1,\ldots,K$}
        \STATE $h_{k} \gets \Gamma_{\theta^{d}_{k}}(h_{k-1}, \mathcal{T}(f_{k-1}), g) $
        \STATE $f_{k} \gets \Lambda_{\theta^{p}_{k}}(f_{k-1}, \mathcal{T}^{*}(h_{k}))$
    \RETURN $f_{K}$
    \ENDFOR
  \end{algorithmic}
\end{algorithm}

\textbf{Forward/backward projection:} For sinogram data, the operator $\mathcal{T}$ and $\mathcal{T^{*}}$ in Algorithm \ref{alg1} is the system matrix $G$ and its adjoint matrix $G^{*}$, which can be precomputed.
%$the forward/backward projection can be done by using the precomputed system matrix $G$ and its adjoint matrix $G^{*}$ multiplied by the data in the image and sinogram domain.
%the forward projection involves multiplying the system matrix $G$ with the image domain data $f_{k}$. On the other hand, the backward projection is done by multiplying the adjoint of $G$, denoted as $G*$, with the sinogram domain data $h_{k}$.
Unlike sinogram data, the corresponding projection matrix $P$ of list-mode data is different for each set of detection data and it cannot be precomputed. This is because \iffalse a row of the projection matrix represents the contribution of each pixel of the image to a LOR, and \fi the sequence of coincidence events in list-mode data is random and disordered, and its length is not fixed. Therefore, for each set of list-mode data, its projection matrix needs to be calculated in real-time before being input to the neural network.

The projection result of a LOR line is the integral of the activity on this line:
\begin{equation}
\label{equ3}
    h(i) = \int_{LOR_{i}}\epsilon f(x,y)ds
\end{equation}
where the $\epsilon$ is the TOF weight, $i\in[1,n]$ represents the i-th LOR, $f(\cdot)$ is the activity distribution and $x\in[1,W], y\in[1,H]$ represents image coordinates. $ds$ represents the line differentiation of the LOR.
\iffalse The value of each element in the projection matrix can be obtained in the process of calculating (\ref{equ3}).\fi
The contribution of each pixel to the LOR varies linearly with the position, so the result of the line integration can be expressed as the product of the length of the line segment and the midpoint activity value of the line segment:

\begin{figure}[t]
    \centering
    \includegraphics[width=8.5cm]{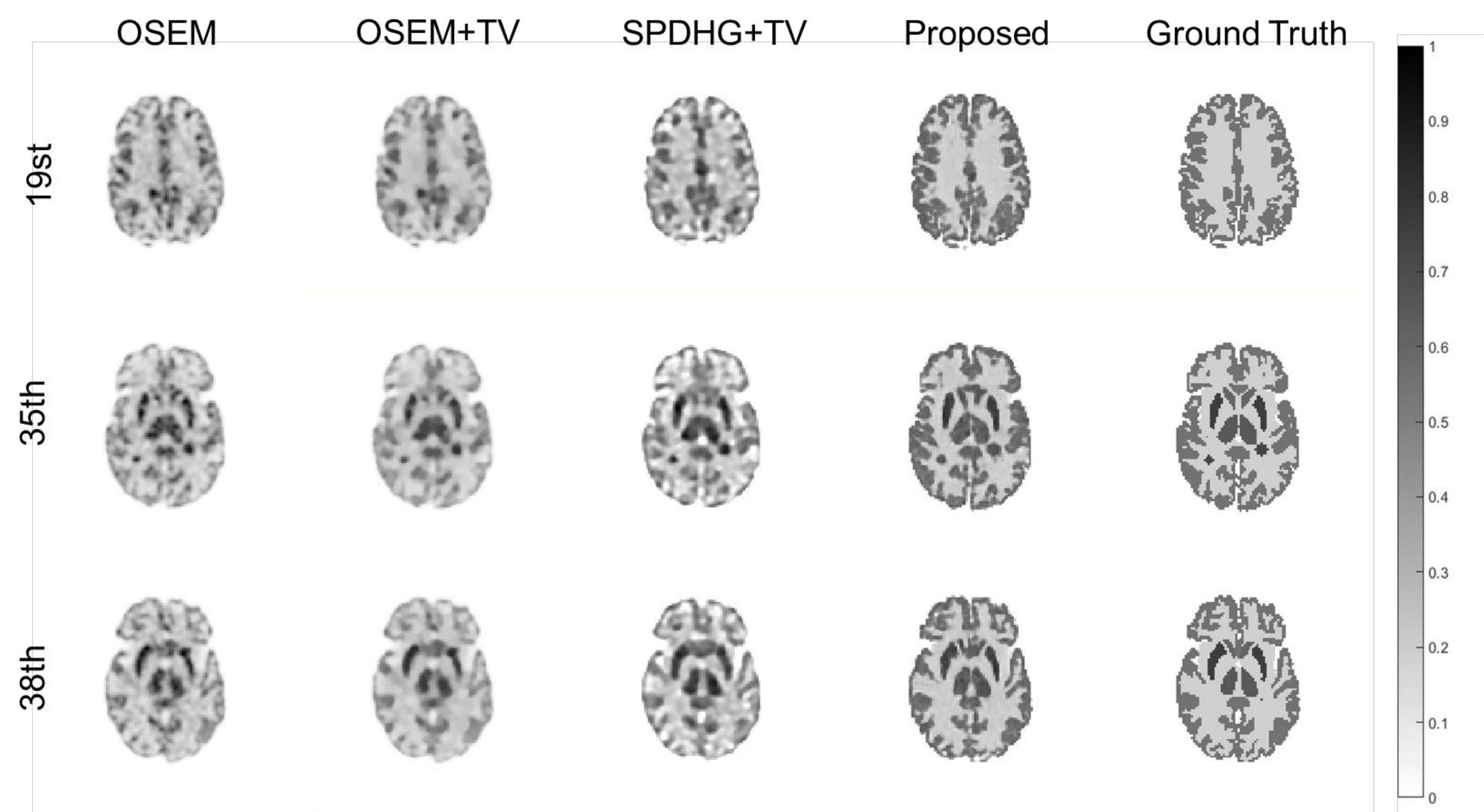}
    \hfill
    \caption{Reconstruction result of OSEM, SPDHG+TV and proposed method. The 19st, 35th, and 38th slices in phantom are selected for display.}
    \label{fig:result1}
\end{figure}

\begin{equation}
\label{equ4}
    h(i) = \sum_{y=1}^{H}\epsilon f(s_{mid}) \cdot\Delta s
\end{equation}
where $\Delta s$ is the length of line segment. The midpoint activity value can be calculated by linear interpolation between two adjacent pixels, then:

\begin{equation}
\label{equ5}
    f(s_{mid}) = \rho f(x_{l}, y) + (1-\rho) f(x_{r}, y)
\end{equation}
$\rho$ is the coefficient of linear interpolation, $x_{l}$ and $x_{r}$ are the coordinates to the left and right of the midpoint. Recall that a single element $P_{ij}$ in the projection matrix $P\in [n, W\times H]$ represents the degree of influence caused by the i-th LOR on the j-th pixel. The result of multiplying $P_{i}$ (the i-th row in $P$) with $f(:)$ (the image resized as a column vector) is $h(i)$. That is:

\begin{equation}
\label{equ6}
    P_{i}\cdot f(:) = \sum_{y=1}^{H}\epsilon[\rho f(x_{l}, y)+(1-\rho)f(x_{r}, y)]\Delta s
\end{equation}
Dividing the terms relating to $f$, then the calculation result of the element value in the projection matrix is:

\begin{equation}
\label{equ7}
    P_{ij} = \epsilon \cdot \rho \cdot \Delta s
\end{equation}

There is no data communication between the calculations of different LOR projections, and the granularity of each LOR projection calculation is relatively fine. Therefore, the projection matrix can be accelerated by CUDA parallel computing. It is worth noting that the projection matrix can be used directly for the subsequent backpropagation of the neural network after it has been computed by requesting memory in the GPU, without the need to copy the data from CPU memory to GPU memory.

\textbf{Dual Module:} The input of the Dual Module is $h_{k}$ and the forward projection result of $f_{k}$. Concatenate two input lists and an all-ones list, and get $\tilde{h} \in \mathbb{R}^{n*3}$. The all-ones list is to serve as the reference, i.e. $g$ in Learned Primal-Dual, for the forward projection result, which is like using the measured original sinogram as the reference during the sinogram projection process. Considering that there is no correlation among the elements of a column in $\tilde{h}$, we need to pay more attention to the connection between the three values of each row. A fully connected (FC) neural network with three input features and one output feature was used, and the dimension of length $n$ becomes the ``channel" dimension. Finally, the output result is used as the residual and the list obtained from the previous iteration $h_{k}$ is summed to obtain the output result of Dual Module.

\textbf{Primal Module:} Primal Modules are similar in structure to Dual modules. The difference is that the input and output of this module are not lists, but images. The input of the Primal Module is the concatenation of $f_{k}$ and the backward projection result of $h_{k+1}$. The Primal module uses Convolutional Neural Networks (CNNs) with PReLU as activation function.

\begin{table}[t]
    \caption{PSNR and SSIM (mean $\pm$ std) of various methods and counts}
    \resizebox{8.7cm}{!}{
    \begin{tabular}{ccccc}
    \toprule
    \multirow{2}*{Methods} & \multicolumn{2}{c}{Counts=3e5} & \multicolumn{2}{c}{Counts=1e5} \\
    \cmidrule(lr){2-3}\cmidrule(lr){4-5}
    & PSNR & SSIM & PSNR & SSIM  \\
    \midrule
    OSEM & 17.68$\pm$1.26 & 0.84$\pm$0.03 & 16.96$\pm$1.22 & 0.82$\pm$0.04 \\
    OSEM+TV & 18.26$\pm$1.25 & 0.85$\pm$0.03 & 17.81$\pm$1.24 & 0.84$\pm$0.03 \\
    SPDHG+TV & 18.37$\pm$1.17 & 0.85$\pm$0.03 & 15.44$\pm$1.10 & 0.80$\pm$0.04 \\
    Proposed & \bf{23.47$\pm$0.49} & \bf{0.94$\pm$0.01} & \bf{21.52$\pm$0.44} & \bf{0.92$\pm$0.01} \\
    \bottomrule
    \end{tabular}
    }
\label{table1}
\end{table}

\begin{table}[b]
    \caption{\label{table2}Comparison of video memory occupation and projection time consuming}
    \resizebox{8.7cm}{!}{
    \begin{tabular}{ccc}
    \toprule %[2pt]
        &  sinogram & list-mode \\
    \midrule %[2pt]
        Memory Occupation & $\sim$87GB & $\sim$20GB \\
        Forward Projection & 364ms & 256ms \\
        Backward Projection & 405ms & 246ms \\
        Compute Projection Matrix & - & 157ms\\
    \bottomrule %[2pt]
    \end{tabular}
    }
\end{table}

\section{EXPERIMENTS}
\label{sec:experiments}

\subsection{Datasets}
\label{ssec:datasets}

% We used the Zubal phantom\cite{zubal1994computerized} with size of 128$\times$128$\times$40 as a starting point for the simulated data. Insert several hot spheres of radius 2mm to 4mm into the phantom at random to simulate the tumor and obtain 12 different phantoms. The compartment model was used along with the kinetic parameters to simulate the activity map. The corresponding sinograms were calculated using the aforementioned forward projection method, and 15$\%$ Poisson random noise was added to sinogram. The time resolution of TOF was set as 400 ps. Different random number seeds were used to generate 12 different Poisson random results for each activity map, resulting in 480 pairs of data. Among them, 400 pairs of data were selected as training set, 40 pairs as validation set, and 40 pairs as test set. Then, the sinograms were scaled down according to the count rate, the maximum count was set to 3e5. Finally, the value in each bin in the sinogram was regarded as the corresponding number of coincident events, and then they were converted into the list-mode data.

The Zubal phantom\cite{zubal1994computerized} with size of 128$\times$128$\times$40 was used as a starting point for the simulated data. We inserted several hot spheres of radius 2mm to 4mm at random positions in phantom to simulate the tumors and obtained 12 different phantoms. Then we got 480 different 2D phantoms of size 128$\times$128. A three compartment model was used along with the Feng's input function to simulate the activity map\cite{feng1993models}. In order to simulate the list-mode data, we first obtained the system matrix for the specific PET scanner using (\ref{equ6}) \iffalse the aforementioned projection method\fi. The system matrix was multiplied by the activity map to obtain the original sinograms. Next, the sinograms were scaled down according to the count rate, the total count was set to 3e5 and 1e5 respectively. Then 15$\%$ Poisson random noise was added to sinograms. Finally, sinograms were converted to list-mode data by rounding down the value of each bin in the sinograms to the number of events in its corresponding position, and using each event as a row in the list-mode data. At this point, we had 480 pairs of activity map and list-mode data. Among them, 400 pairs of data were selected as the training set, 40 pairs as the validation set, and 40 pairs as the test set. 

%It should be noted here that the sinogram data format was used during the simulation of the data, as it is convenient to use sinogram to control the count rate to achieve the right experimental conditions. The sinogram was not required for the subsequent training of the neural network.

The system we used was a simulated classic cylindrical PET scanner, which included 28 modules with 16 crystals each, having a width of 4 mm in the radial direction. Set 17 TOF-bins on each LOR, each TOF-bin is 15mm long and the time resolution of TOF was set as 400 ps. Thus the dimension of the sinogram was 357$\times$224$\times$17. It is worth noting that axial information in the detection system was not required since the experiment focused on 2D reconstruction of PET images.

\subsection{Training Details}
\label{ssec:traininng details}

The number of network iteration layers was set to 7. Each Dual-Net consisted of two hidden layers with 32 features, and Primal-Net consisted of 4 convolutional layers, the number of channels in the first three layers is set to 64, except for the last layer where the number of output channels is 1. The MSE Loss between the output images $f_{n}$ and the ground truth was chosen as the loss function, and the learning rate was set to 1e-5. The training epochs are 500.

\subsection{Results}
\label{ssec:results}

% \begin{table}[t]
%     \caption{\label{table1}PSNR and SSIM (mean $\pm$ std) of various methods}
%     \resizebox{8.5cm}{!}{
%     \begin{tabular}{ccc}
%     \toprule %[2pt]
%         Methods &  PSNR & SSIM \\
%     \midrule %[2pt]
%         MLEM & 18.21$\pm$1.2613 & 0.849$\pm$0.0308 \\
%         SPDHG+TV & 18.19$\pm$1.2112 & 0.850$\pm$0.0300 \\
%         Proposed & \bf{22.83$\pm$ 0.4662} & \bf{0.938$\pm$ 0.0084} \\
%     \bottomrule %[2pt]
%     \end{tabular}
%     }
% \end{table}

To quantitatively evaluate the quality of the reconstruction results, we calculated the Peak Signal-to-Noise Ratio (PSNR) and Structural Similarity Index Measure (SSIM) between the reconstructed images and the ground truth. The comparison methods included the traditional OSEM algorithm, OSEM with TV regularization, and the SPDHG algorithm with TV regularization. The traditional algorithms all had 30 iterations and were divided into 4 subsets. The results are presented in Figure \ref{fig:result1} and Table \ref{table1}. The findings suggest that, for both 3e5 counts and 1e5 counts, the proposed method outperformes the traditional model-based reconstruction algorithm in terms of both quantitative and qualitative analysis. Notably, as no other model-based deep learning method of list-mode reconstruction currently exists, we did not compare our proposed method with other learning-based methods.

In Table \ref{table2}, we conducted a comparison between using sinogram and list-mode data for reconstruction under the experimental settings described above. Specifically, we compared the video memory occupation and the time required for each forward/backward projection. The results indicated that using list-mode data as the starting point resulted in lower memory consumption and less projection time compared to using sinogram data. Due to the need of calculating the projection matrix in real-time, the method based on list-mode data consumed a part of the time when each new set of data was fed into the neural network. However, this time consumption only occurs before the new set of data enters the overall reconstruction process, not before each forward/backward projection. And after the acceleration of parallel computing, this time-consuming was small and negligible.

% However, since the projection matrix needed to be calculated in real-time, a small amount of time was consumed when each new set of data was fed into the neural network. This time-consuming factor was largely mitigated through the acceleration of parallel computing, rendering it negligible. Overall, these findings suggest that using list-mode data as the starting point for reconstruction can offer significant advantages in terms of memory usage and projection time, while also being compatible with real-time processing.

\subsection{Ablation Study}
\label{ssec:ablation study}

\begin{figure}[t]
    \centering
    \includegraphics[width=8.5cm]{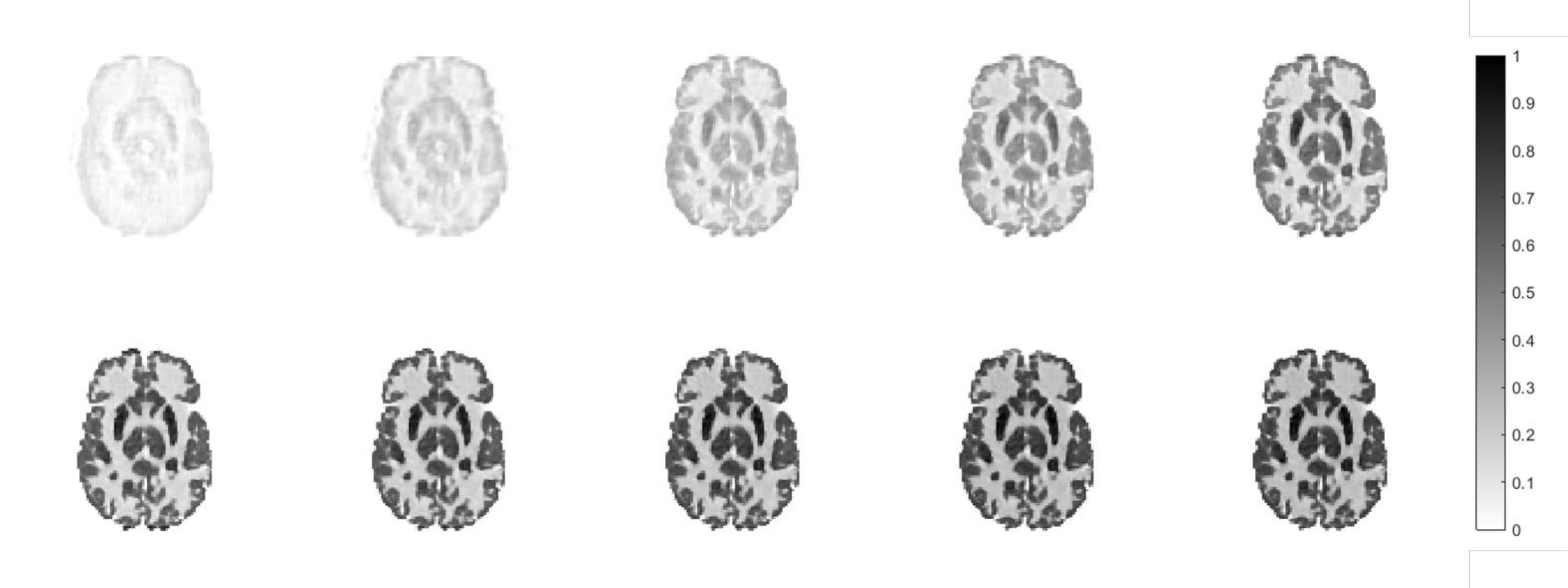}
    \hfill
    \caption{The output of each layer of the proposed method. The first row from left to right represents layers 1 to 5 and the second row from left to right represents layers 6 to 10.}
    \label{fig:result2}
\end{figure}

To balance the effectiveness of the proposed method against its memory usage, we need to investigate the optimal number of iterative layers to be employed. Initially, we set the number of layers to 10, and the output of each layer was examined and compared with the ground truth. The results are presented in Figure \ref{fig:result2}. Our analysis revealed that the output of each layer became increasingly more accurate as the number of iterative layers increased, eventually stabilizing at around the 7th layer. Based on these findings, we set the number of layers to 7 for our experiments.

\section{Conclusion}
\label{sec:conclusion}
This paper proposed a model-based deep learning method for TOF-PET reconstruction from list-mode data called LMPDNet, which was the first application of model-based deep learning to TOF-PET reconstruction.The results showed that the proposed method outperformed traditional iteration-based algorithms in terms of reconstruction quality and outperformed reconstruction from sinogram in terms of spatial and temporal consumption.
%In this paper, we introduced a novel unrolled model-based deep learning method for TOF-PET reconstruction based on list-mode data, named LMPDNet. To the best of our knowledge, this is the first application of model-based deep learning to list-mode reconstruction. Our results indicate that the proposed method outperforms traditional iteration-based algorithms in terms of reconstruction quality, while also exhibiting superior spatial and temporal consumption when compared to reconstruction from sinogram. These findings demonstrate the potential of our approach to improve the accuracy and efficiency of TOF-PET image reconstruction, and highlight its applicability to a range of medical imaging applications.
\bibliographystyle{IEEEbib}
\bibliography{main}

\end{document}